\documentclass[12pt,a4paper]{article}

\usepackage[utf8]{inputenc}

\usepackage{amsmath}
\usepackage{amsfonts}
\usepackage{amssymb}
\usepackage{cite}

\usepackage[retainorgcmds]{IEEEtrantools}

\usepackage{multirow}
\usepackage{multicol}
\usepackage{graphicx}
\usepackage{tikz}
\usepackage{niceframe}

\def\a{{\alpha}}      
\def\b{{\beta}}
\def\g{{\gamma}}
\def\d{{\delta}}
\def\r{{\rho}}
\def\s{{\sigma}}

\def\e{{\epsilon}}

\def\ad{{\dot{\alpha}}}  
\def\bd{{\dot{\beta}}}
\def\gd{{\dot{\gamma}}}

\def\rd{{\dot{\rho}}}

\def\ed{{\bar{\epsilon}}}

\def\ba{{\bar{a}}}       

\def\lvac{{|0\rangle}}
\def\rvac{{|\bar{0}\rangle}}
\def\vac{{|0,\bar{0}\rangle}}
\def\N{{{\bf {\cal N}}}}

\def\z{{\zeta}}

\def\n{{\eta}}

\def\D{{\rm D}}         
\def\Dd{{\bar{\rm D}}}
\def\pa{\partial}

\def\N{{\bar{N}}}
\def\ec{{\e_{\a}{a^\dag}{}^{\a}}}
\def\ea{{\e^{\b}a_{\b}}}
\def\ecd{{\ed_{\ad}{\ba^\dag}{}^{\ad}}}
\def\ead{{\ed^{\bd}\ba_{\bd}}}

\def\[{\left[}
\def\]{\right]}

\def\be{\begin{equation}}
\def\ee{\end{equation}}
\def\bea{\begin{IEEEeqnarray*}}
\def\eea{\end{IEEEeqnarray*}}

\parskip=\medskipamount                 
\thicklines                         

\begin{document}

\title{
{\bf{}BRST Analysis of the Supersymmetric Higher Spin Field
Models}}

\author{\sc I.L. Buchbinder${}^{a}$\thanks{\tt joseph@tspu.edu.ru},
K. Koutrolikos${}^{a,b,c}$\thanks{konstantinos.koutrolikos@gmail.com}
\\[0.4cm]
\it ${}^a$Department of Theoretical Physics,\\
\it Tomsk State Pedagogical University,\\
\it Tomsk 634041, Russia\\
and\\
National Research Tomsk State University\\
Tomsk 634050, Russia\\
[0.26cm]
\it ${}^b$Physics Division, National Technical University of Athens,\\
\it 15780 Zografou Campus, Athens, Greece\\
[0.26cm]
\it \hspace{-13.49pt}${}^{c}$Institute for Theoretical Physics \& Astrophysics, Masaryk University,\\
\it 611 37 Brno, Czech Republic}

\date{}

\maketitle
\thispagestyle{empty}

\begin{abstract}
We develop the BRST approach for all massless integer and
half-integer higher spins in 4D Minkowski space, using the two
component spinor notation and develop the Lagrangian formulation for
supersymmetric higher spin models. It is shown that the problem of
second class constraints disappears and the BRST procedure becomes
much more simple than in tensorial notation. Furthermore, we
demonstrate that the BRST procedure automatically provides extra
auxiliary components that belong in the set of supersymmetry
auxiliary components. Finally, we demonstrate how supersymmetry
transformations\hyphenation{tran-sfo-rma-tions} are realized in such an approach. As a result, we
conclude that the BRST approach to higher spin supersymmetric theories
allows to derive both the Lagrangian and the supersymmetry transformations. Although most part of the work is devoted to massless component supersymmetric
models, we also discuss generalization for massive component
supersymmetric models and for superfield models.

\end{abstract}

\maketitle
\newpage


\section{Introduction}
Supersymmetry is still a very active and promising research area in
modern theoretical physics with many applications in quantum field
theory, mathematical physics, string theory, phenomenology of
elementary particles and cosmology (see e.g. the books \cite{BOOK1},
\cite{BOOK2}, \cite{BOOK3}, \cite{BOOK4} and the references
therein). Another actively developing area of modern theoretical
physics is higher spin field theory (see e.g. the reviews
\cite{Reviews} and the references therein\footnote{A brief review of
the Lagrangian formulation for 4D massless higher spins is
given in \cite{BOOK3} as well.}). In this paper we begin a
systematic formulation of a generic approach to construct the
Lagrangian for supersymmetric higher spin theories. It is clear that
constructing supersymmetric higher spin models, one has to face problems from the
side of supersymmetry and from the side of higher
spins. In particular, we can expect that off-shell Lagrangian
formulation of the supersymmetric higher spin models will require
finding the auxiliary fields both to closure a supersymmetry algebra
off shell and for consistent higher spin field Lagrangian
description.

Four-dimensional ${\cal N}=1$ supersymmetric massless higher spin field
models have been proposed in \cite{curt}.  These models included the
necessary sets of auxiliary fields to provide the Lagrangian
formulation of the higher spin theory but they were on-shell
theories from supersymmetric point of view. It is generally accepted
that the off-shell supersymmetric models can be realized in the
framework of the superfield approach where the set of the auxiliary
fields relevant to off-shell supersymmetry is part of the appropriate
superfield. Massless superfield models in 4D Minkowski space
have been constructed in \cite{Kuz1} and in $AdS_4$
have been found in \cite{Kuz2} (see a brief review in
\cite{BOOK3}). This approach was
further developed in \cite{Kuz3} and some aspect were later studied
in \cite{Konst}.

Lagrangian construction for the massive on-shell 4D, ${\cal N}=1$
higher superspin multiplets in flat space was realized in
\cite{zin1} and for the massive 3D supermultiplets in flat space in
\cite{zin2}\footnote{In this paper the Lagrangian formulation of
3D massless supermultiplets was considered as well.}. Some examples
of the 4D, ${\cal N}=1$ superfield dynamical models in flat
space-time have been constructed in \cite{Buch1}, \cite{Buch2},
\cite{Konst1}. General superfield Lagrangian formulation for
arbitrary massive higher spin supermultiplets is undeveloped until
now.

The generic approach to develop the Lagrangian formulation
for higher spin field theories is realized in the framework of the
BRST-BFV construction \cite{BRST-BFV} (see also the reviews
\cite{Hen}). BRST construction allows us to derive the Lagrangians
for bosonic and fermionic, massless and massive, free and
interacting higher spin fields in the flat and AdS spaces of various
dimensions (see e.g. \cite{Buch3}, \cite{Buch4}, \cite{Buch5},
\cite{different} and the references therein). In this paper we will
focus on the BRST construction for deriving the Lagrangian
formulation of the 4D massless higher spin ${\cal N}=1$
supersymmetric field models.

The first step of the higher spin BRST approach, is to convert all
constraints required for the definition of an irreducible
representations of the Poincare or AdS group, into operators acting
on a Fock space.  These operators should form a first class algebra
in terms of commutators, therefore one can treat them as the
generators of some still unknown gauge theory. The second step is the
construction of the BRST charge on the base of the above constraints.
The third step is deriving the equations of motion as the annihilation of
the physical state by the BRST charge and finding the corresponding
free Lagrangian. Interaction in such a framework means a deformation of the
BRST charge. However, in many cases the above approach faces the
annoying feature of second class constraints, usually related with
the trace and $\gamma$-trace constraints of higher spins. The typical method of
resolving this issue is to expand the Fock space by introducing
extra ghost oscillators which allow the conversion of the second
class constraints to first class and preserve the algebra at the
same time (see e.g. \cite{2nd-1st-class} for conversion procedure in
general gauge theories). In this work, we demonstrate an alternate
method to avoid the second class constraints in the higher spin
field theories by using fields with two-component spinorial indices
instead of vector indices. The advantage of this description is that
the definition of the irreducible representations does not lead to
second class constraints at all. Therefore, in this case the BRST
approach to free, massless, integer and half-integer higher spins
becomes very simple and straight forward.

The motivation for the usage of the spinorial index notations comes
from the desire to apply the BRST approach to 4D ${\cal N}=1$
superspace and describe supersymmetric higher spin theories.  The
natural space were these representation live is the tensor product
of left and right representation of $sl(2,R)$. We will see that the
BRST approach is powerful enough tool for deriving the Lagrangian
description that not only provides the appropriate compensating
fields for the description of higher spin, but also naturally gives
rise to part of the auxiliary states required for by supersymmetry.

Moreover, we demonstrate how supersymmetry transformations emerge in the
Fock space states and we derive the exact on-shell supersymmetry
transformations laws for the components of massless integer and
half-integer superspin irreducible representations.

The paper is organized as follows. In section 2 we briefly review
the constraints required for the description of the irreducible
representations, the Lagrangian formulation and equations of motion
for both integer and half-integer, massless spins. In section 3 we
present the algebra of all operators involved and construct the most
general BRST charge $Q$. In sections 4 and 5 we take limits of $Q$
in order to define $Q_F$ and $Q_B$, the BRST charges that describe
the half-integer and integer irreducible systems. In section 6 we
present, the higher spin supersymmetric Lagrangian and the
supersymmetry transformations between the bosonic and fermionic
theory. In sections 7 and 8 we discuss generalizations for the
massive case and superspace.

\section{Massless Irreducible Representations of the\\ Poincar\'{e} Group}

\subsection{Half-Integer Spin}
The irreducible representations of the Poincar\'{e} group for massless, half-integer spin ($s+1/2$) are defined by the following list of constraints\footnote{The notation $\phi_{\a(k)\ad(l)}$ means that the field $\phi$ has $k$ undotted indices $\a_1\a_2...\a_k$ that are symmetrized and similarly for $l$ dotted indices} :
\bea{l}
\pa^{\b\bd}\psi_{\b\a(s)\bd\ad(s-1)}=0~,~ i\pa^{\b}{}_{(\bd}\psi_{\b\a(s)\ad(s))}=0~,~\Box\psi_{\a(s+1)\ad(s)}=0\IEEEyesnumber
\label{con.}
\eea
In order to generate these constraints from the variation of a Lagrangian, we have to introduce two more
compensators $\bar{\psi}_{\a(s-1)\ad(s)}$ and $\psi_{\a(s-1)\ad(s-2)}$ which also make the Lagrangian to have a gauge symmetry.
This Lagrangian (up to an overall sign) has the following form
\bea{ll}
\mathcal{L}_F=
&~~i\bar{\psi}^{\a(s)\ad(s+1)}\pa^{\a_{s+1}}{}_{\ad_{s+1
}}\psi_{\a(s+1)\ad(s)}\\
&+i\left[\frac{s}{s+1}\right]~\psi^{\a(s+1)\ad(s)}\pa_{\a_{
s+1}\ad_s}\psi_{\a(s)\ad(s-1)}+c.c.\\
&-i\left[\frac{2s+1}{(s+1)^2}\right]~\bar{\psi}^{\a(s-1)\ad(
s)}\pa^{\a_{s}}{}_{\ad_{s}}\psi_{\a(s)\ad(s-1)}\IEEEyesnumber\\
&+i\psi^{\a(s)\ad(s-1)}\pa_{\a_{s}\ad_{s-1}}\psi_{\a(s-1
)\ad(s-2)}+c.c.\\
&-i\bar{\psi}^{\a(s-2)\ad(s-1)}\pa^{\a_{s-1}}{}_{\ad_{s-1
}}\psi_{\a(s-1)\ad(s-2)}
\eea
and is invariant under the transformations.
\bea{l}
\d_G\psi_{\a(s+1)\ad(s)}
=\frac{1}{s!(s+1)!}\pa_{(\a_{s+1}(\ad_s}\lambda_{\a(s))\ad(s-1))}~,\\
\d_G\bar{\psi}_{\a(s-1)\ad(s)}=
\frac{1}{s!}\pa^{\a_{s}}{}_{(\ad_s}\lambda_{\a(s)\ad(s-1)
)}~,\IEEEyesnumber\label{gt-h.i.}\\
\d_G\psi_{\a(s-1)\ad(s-2)}=\frac{s-1}{s}\pa^{\a_s\ad_{
s-1}}\lambda_{\a(s)\ad(s-1)}~. \eea where $\lambda_{\a(s)\ad(s-1)}$
are the unconstrained gauge parameters. The equations of motion
corresponding to the Lagrangian are \bea{l}\IEEEyesnumber\label{eq.}
i\pa^{\a_{s+1}}{}_{(\ad_{s+1}}\psi_{\a(s+1)\ad(s))}-i\frac{s}{(s+1)s!}\pa_{(\a_s(\ad_{s+1}}\bar{\psi}_{\a(s-1))\ad(s))}=0~,\IEEEyessubnumber\\
i\frac{2s+1}{s!}\pa_{(\a_s}{}^{\ad_s}\bar{\psi}_{\a(s-1)\ad(s)}+i{s(s+1)}\pa^{\ad_{s+1}\ad_s}\psi_{\a(s+1)\ad(s)}\\
~~~~~~~~~~~~~~~~~~~~~~~-i\frac{(s+1)^2}{s!(s-1)!}\pa_{(\a_s(\ad_{s-1}}\psi_{\a(s-1))\ad(s-2))}=0~,\IEEEyessubnumber\\
\frac{i}{(s-1)!}\pa^{\a_{s-1}}{}_{(\ad_{s-1}}\psi_{\a(s-1)\ad(s-2))}-i\pa^{\a_{s-1}\ad_{s}}\bar{\psi}_{\a(s-1)\ad(s)}=0~.\IEEEyessubnumber
\eea
It is easy to verify that the constraints (\ref{con.}) can be derived from (\ref{eq.}) once we use the gauge freedom (\ref{gt-h.i.}) to gauge away $\bar{\psi}_{\a(s-1)\ad(s)}$ and $\psi_{\a(s-1)\ad(s-2)}$.
\subsection{Integer spin}
The constraints for the description of a massless, integer spin $(s)$ are
\bea{l}
\pa^{\b\bd}h_{\b\a(s-1)\bd\ad(s-1)}=0~,~\Box h_{\a(s)\ad(s)}=0\IEEEyesnumber
\label{con.B}
\eea
The corresponding Lagrangian includes one real compensator $h_{\a(s-2)\ad(s-2)}$ and has the form
\bea{ll}
\mathcal{L}_B=
&~h^{\a(s)\ad(s)}\Box h_{\a(s)\ad(s)}\\
&-\frac{s}{2}~h^{\a(s)\ad(s)}\pa_{\a_s\ad_s}\pa^{\g\gd}
h_{\g\a(s-1)\gd\ad(s-1)}\\
&+s(s-1)~h^{\a(s)\ad(s)}\pa_{\a_s\ad_s}\pa_{\a_{s-1}
\ad_{s-1}}h_{\a(s-2)\ad(s-2)}\IEEEyesnumber\\
&-s(2s-1)~h^{\a(s-2)\ad(s-2)}\Box h_{\a(s-2)\ad(s-2)}\\
&-\left[\frac{s(s-2)^2}{2}\right]h^{\a(s-2)\ad(s-2)}\pa_{
\a_{s-2}\ad_{s-2}}\pa^{\g\gd}h_{\g\a(s-3)\gd\ad(s-3)}~.
\eea
It is invariant under the following gauge transformations
\bea{l}
\d_G h_{\a(s)\ad(s)}=\frac{1}{s!s!}\pa_{(\a_s(\ad_s}\zeta_{
\a(s-1))\ad(s-1))}~,\IEEEyesnumber\label{gt-i}\\
\d_G h_{\a(s-2)\ad(s-2)}=\frac{s-1}{s^2}\pa^{\a_{s-1}\ad_{
s-1}}\zeta_{\a(s-1)\ad(s-1)}~. \eea where $\zeta_{ \a(s-1)\ad(s-1)}$
are the unconstrained gauge parameters. The equations of motion
corresponding to the Lagrangian are \bea{l}
\IEEEyesnumber\label{eq.B}
\Box h_{\a(s)\ad(s)}-\frac{s}{2 s!^2}\pa_{(\a_s (\ad_s}\pa^{\g\gd}h_{\g\a(s-1))\gd\ad(s-1))}\IEEEyessubnumber\\
~~~~~~~~~~~~~+\frac{s(s-1)}{2 s!^2}\pa_{(\a_s (\ad_s}\pa_{\a_{s-1}\ad_{s-1}}h_{\a(s-2))\ad(s-2))}=0~,\\
\Box h_{\a(s-2)\ad(s-2)}+\frac{(s-2)^2}{2(2s-1) (s-2)!^2}\pa_{(\a_{s-2} (\ad_{s-2}}\pa^{\g\gd}h_{\g\a(s-3))\gd\ad(s-3))}\IEEEyessubnumber\\
~~~~~~~~~~~~~~~~~~-\frac{(s-1)}{2(2s-1)}\pa^{\b\bd}\pa^{\g\gd}h_{\b\g\a(s-2)\bd\gd\ad(s-2)}=0~.
\eea
Using the symmetry freedom (\ref{gt-i}) and the
above equations of motion we can generate the constraints (\ref{con.B}).
\section{Algebra of operators and BRST charge}
First of all we construct the Fock space by introducing two commuting pairs of creation and annihilation operators, one for the left space (undotted indices) and one for the right space (dotted indices).
\bea{llll}
[a_{\a},a_{\b}]=0~,~ & [a_\a , {a^\dag}{}^\b]=\d_{\a}{}^{\b}~,~ & [{a^\dag}{}^{\a},{a^\dag}{}^{\b}]=0~,& \\
{}[\ba_{\ad},\ba_{\bd}]=0~,~ & [\ba_\ad , {\ba^\dag}{}^{\bd}]=\d_{\ad}{}^{\bd}~,~ & [{\ba^\dag}{}^{\ad},{\ba^\dag}{}^{\bd}]=0~,& \IEEEyesnumber\\
{}[a_{\a},\ba_{\bd}]=0~,~ & [a_\a , {\ba^\dag}{}^{\bd}]=0~,~ & [{a^\dag}{}^{\a},{\ba^\dag}{}^{\bd}]=0~,~& [{a^\dag}{}^{\a},\ba_{\bd}]=0~.
\eea

For each one of them, we can define a vacuum state:
\bea{l}
a_{\a}\lvac=0~,~\ba_{\ad}\rvac=0\IEEEyesnumber
\eea
hence, the general vector states are of the form
\bea{l}
|\Phi\rangle=\sum_{k=0}^{\infty}\sum_{l=0}^{\infty}\Phi_{\a(k)\ad(l)}{a^\dag}{}^{\a(k)}{\ba^\dag}{}^{\ad(l)}\vac\IEEEyesnumber
\label{gv}
\eea
and due to the commuting properties of ${a^\dag}{}^{\a}$s and ${\ba^\dag}{}^{\ad}$s the field $\Phi_{\a(k)\ad(l)}$, which is defined as the coefficient in the expansion of the state vector, will have the correct index symmetries for the description of higher spins (see Appendix A).

The set of operators that act upon the general state vectors includes the two number operators (one for left and one for right space)
\bea{l}
N={a^\dag}{}^{\a}a_{\a}~,~\bar{N}={\ba^\dag}{}^{\ad}\ba_{\ad}\IEEEyesnumber\label{0d}
\eea
the derivative operator (there are four combinations)
\bea{ll}
L_{1}=\pa^{\a\ad}a_{\a}\ba_{\ad}~,~L_{-1}=-\pa_{\a\ad}{a^\dag}{}^{\a}{\ba^\dag}{}^{\ad}~,\IEEEyesnumber\label{1d}\\
T_{0}=i\pa^{\a}{}_{\ad}a_{\a}{\ba^\dag}{}^{\ad}~,~T'_{0}=i\pa_{\a}{}^{\ad}{a^\dag}{}^{\a}\ba_{\ad}
\eea
and of course, the $L_0=\Box$ operator.
The coefficients and sings in the definitions ($\ref{1d}$) have been chosen so under complex conjugation we get
$[T_0]^*=T'_0 ~,~[L_1]^*=L_1 ~,~ [L_{-1}]^*=L_{-1}$ and under Hermitian conjugation we get $[T_0]^\dag=T_0 ~,~[T'_0]^\dag=T'_0~,~[L_1]^\dag=L_{-1}$. The algebra of these operators is
\bea{l}
\[L_0 , L_1\]=\[L_0 , L_{-1}\]=\[L_0 , T_0\]=\[L_0 , T'_0\]=\[L_0 , N\]=\[L_0 , \bar{N}\]=0~,\\
\[T_0 , L_1\]=\[T_0 , L_{-1}\]=\[T_0 , L_1\]=\[T'_0 , L_1\]=\[T'_0 , L_{-1}\]=0~,\\
\[L_1,L_{-1}\]=-(N+\bar{N}+2)L_0~,~\[T_0,T'_0\]=(\bar{N}-N)L_0~,\\
T'_0 T_0=L_{-1}L_1 +N(\bar{N}+1)L_0~,~T_0 T'_0=L_{-1}L_1 +(N+1)\N L_0\IEEEyesnumber\label{algebra}\\
\[L_1 , N\] =L_1~,~\[L_{-1}, N\] =-L_{-1}~,~\[T_0, N\] =T_0~,~\[T'_0, N\] =-T'_0~,\\
\[L_1, \N\] =L_1~,~\[L_{-1}, \N\] =-L_{-1}~,~\[T_0, \N\] =-T_0~,~\[T'_0, \N\] =T'_0~.
\eea

Using the above algebra it is possible to construct a nilpotent operator $Q$ ($Q^2=0$)\footnote{In principle nilpotency means that there is a positive number $k$ such that $Q^k=0$. For this case $k$ will be $2$ because of the $g\times g\to g$ structure of the algebra, but in principle for more complicated cases we may have to allow for higher values}. For the case of massless theories (there are no dimension-full parameters) $Q$ must be of the form
\bea{l}
Q=Q^{(0)}+Q^{(1)}+Q^{(2)}\IEEEyesnumber
\eea
where $Q^{(n)}$ includes terms with exactly $n$ derivatives\footnote{For a more abstract discussion $n$ does not have to stop at 2, but since we want to generate equations with up to two derivatives, it is obvious that $n\leq 2$}. Therefore, $Q^{(0)}$ will be a function of $N$ and $\N$
\bea{l}
Q^{(0)}=f(N,\N)\IEEEyesnumber\label{Q0}
\eea
$Q^{(1)}$ will be a linear combination of $L_1,~L_{-1},~T_0,~T'_0$
\bea{l}
Q^{(1)}=\n T_0 +\r T'_0 +\s L_1 +\z L_{-1}\IEEEyesnumber\label{Q1}
\eea
and $Q^{(2)}$ will include $L_0$ plus all possible pair products of the one derivative operators
\bea{l}
Q^{(2)}=\xi L_0 + \phi_1 L_{-1}L_{1}+\phi_2 L_1T_{0}+... \IEEEyesnumber\label{Q2}
\eea

Furthermore due to nilpotency they must be such that
\bea{l}
\[Q^{(0)}\]^2~,~\{Q^{(0)},Q^{(1)}\}=0~,~\{Q^{(0)},Q^{(2)}\}+\[Q^{(1)}\]^2=0~,\IEEEyesnumber\label{npc}\\
\{Q^{(1)},Q^{(2)}\}=0~,~\[Q^{(2)}\]^2=0~.
\eea
It is straight forward to prove that $(\ref{npc})$ are satisfied if
\bea{l}
\phi_1=\phi_2=\dots =0~,\\
\n^2=\r^2=\s^2=\zeta^2=\xi^2=P_{\xi}^2=0~,\\
\{\n,\r\}=\{\s,\zeta\}=\{\n,\s\}=\{\r,\s\}=\{\n,\zeta\}=0~,\IEEEyesnumber\label{gh1}\\
\{\xi,\n\}=\{\xi,\r\}=\{\xi,\s\}=\{\xi,\zeta\}=\{\r,\zeta\}=0~,\\
\{P_{\xi},\n\}=\{P_{\xi},\r\}=\{P_{\xi},\s\}=\{P_{\xi},\zeta\}=0,~\{\xi,P_{\xi}\}=1~,\\
f(N,\N)=\[\n\r(N-\N)+\s\zeta(N+\N+2)\]P_{\xi}~.
\eea

We conclude that
\begin{enumerate}
\item Based on $(\ref{gh1})$ the objects $\n,~\r~,\s~,\zeta,~\xi$ must be interpreted as anti-commuting oscillators that expand our Fock space. These are the usual BRST ghosts and we assign them a ghost number $+1$.
\item For each one of them, there is a conjugate oscillator $P_{\n},~P_{\r},~P_{\s},~P_{\zeta},~P_{\xi}$ which has a ghost number of $-1$.
\item We can define a vacuum state $|0_{gh}\rangle$ for the ghost sector, as the state that is being annihilated by a subset of the above oscillators. Different choices for the vacuum state (or equivalently different choices of oscillators that annihilate the vacuum) will lead to the description of different systems.
\item The BRST charge $Q$ for algebra $(\ref{algebra})$ is:
\bea{l}
\hspace{-2em}Q=\n T_0 +\r T'_0 +\s L_1 +\z L_{-1}+\xi L_0+\[\n\r(N-\N)+\s\zeta(N+\N+2)\]P_{\xi}\IEEEyesnumber\label{gBRST}
\eea
\item By choosing $\s$ to be the hermitian conjugate of $\z$ and $\n,~\r,~\xi$ to be self hermitian then $Q$ becomes hermitian.
\end{enumerate}

For both integer and half-integer spins $T'_0,~L_{-1}$ are not constraints, therefore a reasonable choice for the vacuum state is
\bea{l}
P_{\n}|0_{gh}\rangle=\r|0_{gh}\rangle=P_{\s}|0_{gh}\rangle=\z|0_{gh}\rangle=P_{\xi}|0_{gh}\rangle=0~.\IEEEyesnumber
\eea
With this choice, the most general state $|\Phi\rangle$ allowed, is of the form
\bea{l}
|\Phi\rangle=\sum_{\substack{k_1,k_2,k_3\\ l_1,l_2}}\n^{k_1}\s^{k_2}\xi^{k_3}P_{\z}{}^{l_1}P_{\xi}^{l_2}|\Phi_{k_1,k_2,K_3,l_1,l_2}\rangle\IEEEyesnumber
\eea
where the $k$s and the $l$s can take two values, zero or one and we sum over them.
The ghost number value for the general term in the sum is (we choose the vacuum to have zero ghost number)
\bea{l}
gh(|\Phi\rangle)=k_1+k_2+k_3-l_1-l_2~.\IEEEyesnumber
\eea
Therefore the zero ghost state, which will play the role of the physical state\\ ($Q|\Psi\rangle=0$) is
\bea{rl}
|\Psi\rangle=|S\rangle+&\n P_{\z}|A\rangle+\s P_{\z}|B\rangle+\xi P_{\z}|\Gamma\rangle\\
+&\n P_{\r}|U\rangle+\s P_{\r}|V\rangle+\xi P_{\r}|\Delta\rangle\IEEEyesnumber\label{0gh}\\
+&\n\s P_{\z}P_{\r}|W\rangle+\n\xi P_{\z}P_{\r}|Z\rangle+\s\xi P_{\z}P_{\r}|H\rangle
\eea
the -1 ghost state, which will play the role of the gauge parameter for the transformation of the physical state ($\delta_{G}|\Psi\rangle=Q|\Lambda\rangle$) is
\bea{rl}
|\Lambda\rangle=P_{\z}|\lambda\rangle + P_{\r}|\kappa\rangle + \n P_{\z}P_{\r}|\pi\rangle+\s P_{\z}P_{\r}|\tau\rangle+\xi P_{\z}P_{\r}|\Upsilon\rangle\IEEEyesnumber\label{-1gh}
\eea
and the -2 ghost state which will play the role of a gauge parameter for a second level gauge transformation ($\delta_{G}|\Lambda\rangle=Q|\Xi\rangle$) is
\bea{rl}
|\Xi\rangle=P_{\z}P_{\r}|\omega\rangle\IEEEyesnumber\label{-2gh}
\eea

Now we can find the equations of motion for the components as defined above:
\bea{l}\IEEEyesnumber\label{e.o.m}
T_0|S\rangle-L_{-1}|A\rangle-T'_{0}|U\rangle+(N-\N)|\Delta\rangle=0~,\IEEEyessubnumber\\
T'_0|V\rangle-L_{1}|S\rangle+L_{-1}|B\rangle+(N+\N+2)|\Gamma\rangle=0~,\IEEEyessubnumber\\
T_0|B\rangle-L_{1}|A\rangle-T'_{0}|W\rangle-(N-\N)|H\rangle=0~,\IEEEyessubnumber\\
T_0|V\rangle-L_{1}|U\rangle+L_{-1}|W\rangle-(N+\N+2)|Z\rangle=0~,\IEEEyessubnumber\\
L_0|S\rangle-L_{-1}|\Gamma\rangle-T'_{0}|\Delta\rangle=0~,\IEEEyessubnumber\\
L_0|A\rangle-T_{0}|\Gamma\rangle+T'_{0}|Z\rangle=0~,\IEEEyessubnumber\\
L_0|B\rangle-L_{1}|\Gamma\rangle+T'_{0}|H\rangle=0~,\IEEEyessubnumber\\
L_0|U\rangle-L_{-1}|Z\rangle-T_{0}|\Delta\rangle=0~,\IEEEyessubnumber\\
L_0|V\rangle-L_{1}|\Delta\rangle-L_{-1}|H\rangle=0~,\IEEEyessubnumber\\
L_0|W\rangle-L_{1}|Z\rangle+T_{0}|H\rangle=0~,\IEEEyessubnumber
\eea
and they are invariant under the following transformations
\bea{ll}
\delta_G |S\rangle=L_{-1}|\lambda\rangle+T'_{0}|\kappa\rangle~,&~\delta_G |V\rangle=L_{1}|\kappa\rangle-L_{-1}|\tau\rangle+\left(N+\N+2\right)|\Upsilon\rangle,\\
\delta_G |A\rangle=T_{0}|\lambda\rangle+T'_{0}|\pi\rangle-\left(N-\N\right)|\Upsilon\rangle~,&~\delta_G |\Delta\rangle=L_{0}|\kappa\rangle-L_{-1}|\Upsilon\rangle,\\
\delta_G |B\rangle=L_{1}|\lambda\rangle+T'_{0}|\tau\rangle~,&~\delta_G |W\rangle=T_{0}|\tau\rangle-L_{1}|\pi\rangle,\IEEEyesnumber\label{gt1}\\
\delta_G |\Gamma\rangle=L_{0}|\lambda\rangle+T'_{0}|\Upsilon\rangle~,&~\delta_G |Z\rangle=-L_{0}|\pi\rangle+T_{0}|\Upsilon\rangle,\\
\delta_G |U\rangle=T_{0}|\kappa\rangle-L_{-1}|\pi\rangle~,&~\delta_G |H\rangle=-L_{0}|\tau\rangle+L_{1}|\Upsilon\rangle,
\eea
\bea{ll}
\delta_G |\lambda\rangle=-T'_{0}|\omega\rangle~,&~\delta_G |\tau\rangle=L_{1}|\omega\rangle,\\
\delta_G |\kappa\rangle=L_{-1}|\omega\rangle~,&~\delta_G |\Upsilon\rangle=L_{0}|\omega\rangle,\IEEEyesnumber\label{gt2}\\
\delta_G |\pi\rangle=T_{0}|\omega\rangle~.&~
\eea

It is obvious, that the above construction does not describe an irreducible representation but gives the most general BRST charge that can be constructed out of the specific set of operators. Therefore, the BRST charges responsible for the irreducible representations must be able to be derived out of it.
We observe that the algebra ($\ref{algebra}$) has three subalgebras
\begin{enumerate}
\item $\{T_0~,L_1~,L_{-1}~,L_0~,N~,\N\}$
\item $\{T'_0~,L_1~,L_{-1}~,L_0~,N~,\N\}$
\item $\{L_1~,L_{-1}~,L_0~,N~,\N\}$
\end{enumerate}
The first one includes the constraints required for the description of massless half-integer spins. The second set is related to the first one via a complex conjugation. Therefore, it describes the same representation as seen from the complex conjugated viewpoint. Finally, the third subalgebra includes the constraints for the description of massless integer spins. In this case all the operators involved are real therefore we can have a real representation.
\section{BRST description for half-integer spins}
To construction the BRST charge generated by the subalgebra of\\
$\{T_0,~L_1,~L_{-1},~L_0,~N,~\N\}$
all we have to do is start with the general BRST charge ($\ref{gBRST}$) and freeze out the $\r,~P_{\r}$ pair of oscillators from the Fock space. We can do that by ignoring all terms in $Q$ that include $\r$ (naive $\r\to 0$ limit). Hence, the BRST charge for the description of massless, half-integer spin, fermions is
\bea{l}
Q_{F}=\n T_0 +\s L_1 +\z L_{-1}+\xi L_0+\s\zeta(N+\N+2)P_{\xi}~.\IEEEyesnumber\label{BRST-F}
\eea
By taking the same limit in equations ($\ref{0gh},\ref{-1gh},\ref{-2gh},\ref{e.o.m}$) we find the physical state (zero ghost state) to be
\bea{rl}
|\Psi_{F}\rangle=|S_{F}\rangle+&\n P_{\z}|A_{F}\rangle+\s P_{\z}|B_{F}\rangle+\xi P_{\z}|\Gamma_{F}\rangle\IEEEyesnumber\label{0gh-F}
\eea
the gauge parameter state (-1 ghost state)
\bea{rl}
|\Lambda_{F}\rangle=P_{\z}|\lambda_{F}\rangle\IEEEyesnumber\label{-1gh-F}
\eea
and there is no second level gauge parameter (-2 ghost state).
The equations of motion are
\bea{l}\IEEEyesnumber\label{e.o.m-F}
T_0|S_{F}\rangle-L_{-1}|A_{F}\rangle=0\IEEEyessubnumber\\
L_{1}|S_{F}\rangle-L_{-1}|B_{F}\rangle+(N+\N+2)|\Gamma_F\rangle=0\IEEEyessubnumber\label{Gamma}\\
T_0|B_{F}\rangle-L_{1}|A_{F}\rangle=0\IEEEyessubnumber\\
L_0|S_{F}\rangle-L_{-1}|\Gamma_{F}\rangle=0\IEEEyessubnumber\\
L_0|A_{F}\rangle-T_{0}|\Gamma_{F}\rangle=0\IEEEyessubnumber\\
L_0|B_{F}\rangle-L_{1}|\Gamma_{F}\rangle=0\IEEEyessubnumber
\eea
and they are invariant under the following transformations
\bea{l}
\delta_G |S_{F}\rangle=L_{-1}|\lambda_{F}\rangle~,~\delta_G |B_{F}\rangle=L_{1}|\lambda_{F}\rangle~,\IEEEyesnumber\label{gt-F}\\
\delta_G |A_{F}\rangle=T_{0}|\lambda_{F}\rangle~,~~~\delta_G |\Gamma_{F}\rangle=L_{0}|\lambda_{F}\rangle~.
\eea

At this point, it is obvious that this system indeed describes a massless half-integer irreducible representation, because the gauge transformations of the three components $|S_{F}\rangle,~|A_{F}\rangle,~|B_{F}\rangle$ match gauge transformations ($\ref{gt-h.i.}$) of the three components required for the description of the irreducible representation and the forth component $|\Gamma_{F}\rangle$ is auxiliary since its equation of motion ($\ref{Gamma}$) is algebraic. As a consequence the dynamics of the two systems must be the same.

However, we can demonstrate that further. $|\Gamma_{F}\rangle$ component is auxiliary due to the fact that it has higher mass dimension, because the oscillator $\xi$ is the coefficient of the $\Box$ operator in the BRST charge $Q_{F}$. Therefore, we are allowed to do a redefinition of it in terms of the other states $|S_{F}\rangle,~|A_{F}\rangle,~|B_{F}\rangle$. Of course, this redefinition must be compatible with the gauge transformation of $|\Gamma_{F}\rangle$. Not only that, but we should be able to use this redefinition to make $|\Gamma_{F}\rangle$ 1) vanish on-shell and 2) make it gauge invariant. With that in mind, consider the ansatz
\bea{l}
|\Gamma_{F}\rangle=g_1 T'_0 |A_{F}\rangle+g_2 L_1|S_{F}\rangle+g_3 L_{-1}|B_{F}\rangle+|\beta\rangle\IEEEyesnumber
\eea
In order the left hand side to be compatible with the right hand side and our demand for $\delta_G|\beta\rangle=0$,
we must have
\bea{l}
g_1+g_2+g_3=0~,~
g_1N(\N+1)-g_2(N+\N+2)=1~.\IEEEyesnumber
\eea
The solution of the above gives the redefinition
\bea{rl}
|\Gamma_{F}\rangle=&\frac{g}{N+\N+2}\Big\{(N+\N+2)T'_0 |A_{F}\rangle+N(\N+1)L_1|S_{F}\rangle\\
&~~~~~~~~~~~~~~~~-(N+1)(\N+2)L_{-1}|B_{F}\rangle\Big\}\\
&-\frac{1}{N+\N+2}\Big\{L_1|S_{F}\rangle-L_{-1}|B_{F}\rangle\Big\}+|\beta\rangle\IEEEyesnumber\label{rdftn-Gamma-F} ~~,~~~~~~~~\forall g
\eea
and the substitution of this expression to ($\ref{e.o.m-F}$) gives the following equations
\bea{l}
|\mathcal{E}_1\rangle=T_0|S_{F}\rangle-L_{-1}|A_{F}\rangle=0~,\\
|\mathcal{E}_2\rangle=(N+\N+2)T'_0|A_{F}\rangle+N(\N+1)L_{1}|S_{F}\rangle-(N+1)(\N+2)L_{-1}|B_{F}\rangle=0~,\IEEEyesnumber\label{BRSTeom-F}\\
|\mathcal{E}_3\rangle=T_0|B_{F}\rangle-L_{1}|A_{F}\rangle=0~,\\
\hline\\
|\mathcal{E}_4\rangle=L_0|S_{F}\rangle+\frac{1}{N+\N}L_{-1}L_1|S_{F}\rangle-\frac{1}{N+\N}L_{-1}L_{-1}|B_{F}\rangle=0~,\IEEEyesnumber\label{BRSTeom2d-F}\\
|\mathcal{E}_5\rangle=\frac{2(N+\N+3)}{N+\N+4}L_0|B_{F}\rangle-\frac{1}{N+\N+4}L_{-1}L_1|B_{F}\rangle+\frac{1}{N+\N+4}L_{1}L_1|S_{F}\rangle=0~,\\
\hline\\
|\mathcal{E}_6\rangle=|\beta\rangle=0~.\IEEEyesnumber\label{BRSTaux-F}
\eea
Equations ($\ref{BRSTeom-F}$) are exactly the equations of motion for a massless half-integer spin ($\ref{eq.}$). Equations ($\ref{BRSTeom2d-F}$) are quadratic in derivatives and have the structure of integer spin equations ($\ref{eq.B}$). However, these equations are not extra constraints because they are automatically satisfied if $|\mathcal{E}_1\rangle,~|\mathcal{E}_2\rangle,~|\mathcal{E}_3\rangle$ hold. This is due to the following identities
\bea{l}\IEEEyesnumber\label{connection}
T'_0 |\mathcal{E}_1\rangle+\frac{1}{N+\N}L_{-1}|\mathcal{E}_3\rangle=N(\N+1)|\mathcal{E}_4\rangle~,\IEEEyessubnumber\\
T'_0 |\mathcal{E}_2\rangle+\frac{1}{N+\N+4}L_{1}|\mathcal{E}_3\rangle=(N+1)(\N+2)|\mathcal{E}_5\rangle~.\IEEEyessubnumber
\eea
Equation ($\ref{BRSTaux-F}$) is the statement that the theory has a gauge invariant auxiliary component. Exactly this type of components appear in the off-shell supersymmetric spectrum of the irreducible representations of the super-Poincar\'{e} group and it is very interesting that BRST is powerful enough to signal about their presence.

Equations ($\ref{BRSTeom-F}, \ref{BRSTaux-F}$) and  can be deduced from the Lagrangian\footnote{up to an overall factor}
\bea{ll}
\mathcal{L}=&~~\langle S_F|T_0|S_F\rangle\\
&-\langle S_F|L_{-1}|A_F\rangle -\langle A_F|L_1|S_F\rangle\\
&-\langle A_F|\frac{N+\N+2}{N(\N+1)}T'_0|A_F\rangle\IEEEyesnumber\label{LF}\\
&+\langle A_F|\frac{(N+1)(\N+2)}{N(\N+1)}L_{-1}|B_F\rangle+\langle B_F|\frac{(N+3)(\N+2)}{(N+2)(\N+1)}L_1|A_F\rangle\\
&-\langle B_F|\frac{(N+3)(\N+2)}{(N+2)(\N+1)}T_0|B_F\rangle\\
&+\langle \r|\beta\rangle+\langle \beta|\r\rangle
\eea
where $|\r\rangle$ is a lagrange multiplier that is required in order to generate the algebraic equation of $|\b\rangle$, since the mass dimensions of $|\b\rangle$ do not allow it to appear in a quadratic way in the lagrangian.  A very intriguing observation is that the above Lagrangian corresponds to the fermionic part of the superspace action that describes massless integer and half-integer superspin theories.
The fields $\psi_{\a(s+1)\ad(s)},~\bar{\psi}_{\a(s-1)\ad(s)},$ $~\psi_{\a(s-1)\ad(s-2)}$ in ($\ref{eq.}$) correspond to the states $|S_{F}\rangle$, $|A_{F}\rangle$, $|B_{F}\rangle$ in ($\ref{BRSTeom-F}$) and the auxiliary states $|\b\rangle,~|\r\rangle$ will correspond to the auxiliary field $\b_{\a(s)\ad(s-1)},~\r_{\a(s)\ad(s-1)}$ respectively.

\section{BRST description of integer spin}
For the massless integer spin representations, we focus at the subalgebra generated by the set
$\{L_1,~L_{-1},~L_0,~N,~\N\}$. In this case we have to freeze both $\n$ and $\rho$ oscillators. Hence, the expression for the integer spin BRST charge, $Q_B$
is
\bea{l}
Q_{B}=\s L_1 +\z L_{-1}+\xi L_0+\s\zeta(N+\N+2)P_{\xi}~.\IEEEyesnumber\label{BRST-B}
\eea
The physical state (zero ghost state) is
\bea{l}
|\Psi_{B}\rangle=|S_{B}\rangle+\s P_{\z}|B_{B}\rangle+\xi P_{\z}|\Gamma_{B}\rangle\IEEEyesnumber\label{0gh-B}
\eea
and the gauge parameter state (-1 ghost state) is
\bea{l}
|\Lambda_{B}\rangle=P_{\z}|\lambda_{B}\rangle~.\IEEEyesnumber\label{-1gh-B}
\eea
Immediatly we get the following equation of motion
\bea{l}\IEEEyesnumber\label{e.o.m-B}
L_{1}|S_{B}\rangle-L_{-1}|B_{B}\rangle+(N+\N+2)|\Gamma_{B}\rangle=0~,\IEEEyessubnumber\\
L_0|S_{B}\rangle-L_{-1}|\Gamma_{B}\rangle=0~,\IEEEyessubnumber\\
L_0|B_{B}\rangle-L_{1}|\Gamma_{B}\rangle=0~.\IEEEyessubnumber
\eea
which are invariant under the following gauge transformations
\bea{l}
\delta_G |S_{B}\rangle=L_{-1}|\lambda_{B}\rangle~,~\delta_G |\Gamma_{B}\rangle=L_{0}|\lambda_{B}\rangle~,\IEEEyesnumber\label{gt-B}\\
\delta_G |B_{B}\rangle=L_{1}|\lambda_{B}\rangle~.
\eea

As in the previous case, the state $|\Gamma_{B}\rangle$ is auxiliary and therefore can be redefined in order to make it vanish on-shell and gauge invariant at the same time. Since there is no state $|A\rangle$ this time, we can get the expression for the redefinition of $|\Gamma_{B}\rangle$ by taking the $g\to$ 0 limit of ($\ref{rdftn-Gamma-F}$)
\bea{l}
|\Gamma_{B}\rangle=-\frac{1}{N+\N+2}\Big\{L_1|S_{B}\rangle-L_{-1}|B_{B}\rangle\Big\}+|\gamma\rangle~.\IEEEyesnumber\label{rdftn-Gamma-B}
\eea
Plugging in this redefinition into ($\ref{e.o.m-B}$) we get the equation of motion
\bea{l}
|\mathcal{Z}_1\rangle=L_0|S_{B}\rangle+\frac{1}{N+\N}L_{-1}L_1|S_{B}\rangle-\frac{1}{N+\N}L_{-1}L_{-1}|B_{B}\rangle=0~,\IEEEyesnumber\label{BRSTeom2d-B}\\
|\mathcal{Z}_2\rangle=\frac{2(N+\N+3)}{N+\N+4}L_0|B_{B}\rangle-\frac{1}{N+\N+4}L_{-1}L_1|B_{B}\rangle+\frac{1}{N+\N+4}L_{1}L_1|S_{B}\rangle=0~,\\
\hline\\
|\mathcal{Z}_3\rangle=|\gamma\rangle=0~.\IEEEyesnumber\label{BRSTaux-B}
\eea

Equations ($\ref{BRSTeom2d-B}$) match exactly the equation for the description of massless integer spin ($\ref{eq.B}$) and as in the previous case, equation ($\ref{BRSTaux-B}$) declares that the theory has a bosonic gauge invariant, auxiliary state. The corresponding Lagrangian is
\bea{ll}
\mathcal{L}=&\langle S_B|L_0|S_B\rangle-\langle S_B|\frac{1}{N+\N}L_{-1}L_1|S_B\rangle\\
&-\langle S_B|\frac{1}{N+\N}L_{-1}L_{-1}|B_B\rangle-\langle B_B|\frac{1}{N+\N+4}L_1L_1|S_B\rangle\IEEEyesnumber\label{Lb}\\
&-2\langle B_B|\frac{N+\N+3}{N+\N+4}L_0|B_B\rangle+\langle B_B|\frac{1}{N+\N+4}L_{-1}L_1|B_B\rangle\\
&+\langle \gamma|\gamma\rangle~.
\eea
The fields $h_{\a(s)\ad(s)},~h_{\a(s-2)\ad(s-2)}$ correspond to the states $|S_{B}\rangle,~|B_{B}\rangle$ and state $|\gamma\rangle$ will correspond to one of the  supersymmetry auxiliary component. Unlike the previous case where we got the full spectrum of supersymmetric auxiliary components, this is no longer true in this case. In principle, we can allow $|\gamma\rangle$ to acquire some internal structure that can reveal the full set of bosonic auxiliary components, but at this approximation we can see them. That can happen if we demand from the very beginning for the theory to be off-shell supersymmetric and encode this information to the BRST charge by introducing the corresponding constraints that relate the bosonic and fermionic degrees of freedom. Of course, this will be automatic if we manage to repeat these type of construction in superspace. Finally, the most traditional way to detect the missing bosonic auxiliary components is to talk about the closure of the on-shell supersymmetry transformations.


\section{Supersymmetric Invariant Theory}
In the previous two sections we have derived the spinorial BRST
description of free massless higher spins. We noticed that the field
that describes the $s+1/2$ spin satisfies the equations of motion of
the field that describes the $s$ spin. Of course this was to be
expected because the constraints that describe the integer spin $s$
are subset of the constraints that describe the half-integer spin
$s+1/2$, but nevertheless it provides some reasoning to why one
should expect the combination of these two systems to have
supersymmetry. Also we have seen that the BRST approach is powerful
enough, that even though we have not mentioned anything about
supersymmetry, it naturally gives rise to components that will play
the role of supersymmetric auxiliary fields, once we introduce
supersymmetry.

\subsection{On-shell formulation}
The above observations, force us to discuss the supersymmetric theory. In this part, we will focus on the on-shell discussion. For that we consider the lagrangian
\bea{l}
\mathcal{L}_S=\mathcal{L}_B|+\mathcal{L}_F|\IEEEyesnumber
\eea
where by $\mathcal{L}_B|$ we mean the bosonic lagrangian ($\ref{Lb}$) with the auxiliary component ($|\gamma\rangle$) integrated out. Similarly, $\mathcal{L}_F|$ is the fermionic lagrangian ($\ref{LF}$) without components $|\b\rangle~,|\rho\rangle$.  It would be interesting to see how this formulation
accommodates on-shell supersymmetry.

First of all, since we have to be able to map bosons to fermions and backwards, we must be able to change the parity of a state $|\Phi\rangle=\Phi_{\a(k)\ad(l)}{a^\dag}{}^{\a(k)}{\ba^\dag}{}^{\ad(l)}|0,\bar{0}\rangle$. The parity of such a state is controlled by the total number of creation operators (free indices) it includes. States with $N+\N$ = even, have zero parity and states with $N+\N$ = odd, have parity one\footnote{A definition of parity in this context can be $\varepsilon=N+\N$ (mod $2$)}. Therefore, to convert a state from one parity to the other we have to change the value of $N+\N$ by a step of one. There is no object that can do this, therefore we introduce a parameter with an undotted index $\e_{\a}$ ($\e_{\a}{a^\dag}{}^{\a},~\e^{\b}a_{\b}$) and its complex conjugate $\ed_{\ad}$ ($\ed_{\ad}{\ba^\dag}{}^{\ad},~\ed^{\bd}\ba_{\bd}$).

For the case of global supersymmetry these parameters are constants and we can prove that they satisfy the following
identities
\bea{ll}
L_1\ec=(N+2)\ec L_1+\frac{i}{N+1}\ea T'_{0}~,~& L_{-1}\ec=\ec L_{-1}~,\\
T_0\ec=(N+2)\ec T_0+\frac{i}{N+1}\ea L_{-1}~,~& T'_{0}\ec=\ec T'_0~,\IEEEyesnumber\label{e-id}\\
L_{-1}\ea=\frac{N}{(N+1)^2}\ea L_{-1}+\frac{i}{N+1}\ec T_{0}~,~& L_{1}\ea=\ea L_{1}~,\\
T'_0\ea=\frac{N}{(N+1)^2}\ea T'_0+\frac{i}{N+1}\ec L_{1}~,~& T_{0}\ea=\ea T_0\\
\hline\\
L_1\ecd=(\N+2)\ecd L_1+\frac{i}{\N+1}\ead T_{0}~,~& L_{-1}\ecd=\ecd L_{-1}~,\\
T'_0\ecd=(\N+2)\ecd T'_0+\frac{i}{\N+1}\ead L_{-1}~,~& T_{0}\ecd=\ecd T_0~,\IEEEyesnumber\label{ed-id}\\
L_{-1}\ead=\frac{\N}{(\N+1)^2}\ead L_{-1}+\frac{i}{\N+1}\ecd T'_{0}~,~& L_{1}\ead=\ead L_{1}~,\\
T_0\ead=\frac{\N}{(\N+1)^2}\ead T_0+\frac{i}{\N+1}\ecd L_{1}~,~& T'_{0}\ead=\ead T'_0~.
\eea

Now using ($\ref{e-id}$) with the fermionic equations ($\ref{BRSTeom-F}$) we can prove the following
\bea{l}\IEEEyesnumber\label{susy1}
|\tilde{\mathcal{E}}_1\rangle\equiv\ea|\mathcal{E}_1\rangle=T_0|\tilde{S}_{B}\rangle-L_{-1}|\tilde{A}_{B}\rangle~,\IEEEyessubnumber\\
|\tilde{\mathcal{E}}_2\rangle\equiv\frac{N}{N+1}\ea|\mathcal{E}_2\rangle-i(N+2)(\N+2)\ec|\mathcal{E}_3\rangle\\
~~~~~=(N+\N+2)T'_0|\tilde{A}_{B}\rangle+N(\N+1)L_1|\tilde{S}_{B}\rangle-(N+1)(\N+2)L_{-1}|\tilde{B}_{B}\rangle,\IEEEyessubnumber\\
|\tilde{\mathcal{E}}_3\rangle\equiv(N+4)\ea|\mathcal{E}_3\rangle=T_0|\tilde{B}_{B}\rangle-L_{1}|\tilde{A}_{B}\rangle\IEEEyessubnumber
\eea
with
\bea{l}
|\tilde{S}_{B}\rangle=\ea|S_{F}\rangle+\frac{i}{N}\ec|A_{F}\rangle~,\\
|\tilde{A}_{B}\rangle=(N+3)\ea|A_{F}\rangle~,\IEEEyesnumber\\
|\tilde{B}_{B}\rangle=(N+3)\ea|B_{F}\rangle~.
\eea
The conclusion is that, the tilded equations are satisfied once the untilded equations are satisfied and also both sets have exactly the same structure. Therefore, states $|\tilde{S}_{B}\rangle,|\tilde{A}_{B}\rangle,|\tilde{B}_{B}\rangle$ on-shell satisfy the same equations as in ($\ref{BRSTeom-F}$) and therefore due to ($\ref{connection}$), the states $|\tilde{S}_{B}\rangle,|\tilde{B}_{B}\rangle$ satisfy exactly the integer spin equations ($\ref{BRSTeom2d-B}$). Hence, we can not distinguish between $(|S_{B}\rangle,|B_{B}\rangle)$ and
$(|S_{B}\rangle+|\tilde{S}_{B}\rangle+c.c.,|B_{B}\rangle+|\tilde{B}_{B}\rangle+c.c.)$.
Therefore, the integer spin theory is invariant under the transformation
\bea{l}
\d_S |S_B\rangle=\ea|S_{F}\rangle+\frac{i}{N}\ec|A_{F}\rangle+c.c.~,\IEEEyesnumber\label{ds1}\\
\d_S |B_B\rangle=(N+3)\ea|B_{F}\rangle+c.c.~.
\eea

Furthermore, if we start from the bosonic equations ($\ref{BRSTeom2d-B}$) and use ($\ref{ed-id}$) we can show that,
\bea{l}\IEEEyesnumber\label{susy2}
|\tilde{\mathcal{Z}}_1\rangle\equiv\ecd|\mathcal{Z}_1\rangle=T_0|\tilde{S}_{F}\rangle-L_{-1}|\tilde{A}_{F}\rangle~,\IEEEyessubnumber\\
|\tilde{\mathcal{Z}}_2\rangle\equiv\frac{(N+2)(\N+1)}{N+3}\ecd|\mathcal{Z}_2\rangle=T_0|\tilde{B}_{F}\rangle-L_1|\tilde{A}_{F}\rangle~,\IEEEyessubnumber\\
|\tilde{\mathcal{Z}}_3\rangle\equiv iN\ead|\mathcal{Z}_1\rangle\\
~~~~~=(N+\N+2)T'_0|\tilde{A}_{F}\rangle+N(\N+1)L_{1}|\tilde{S}_{F}\rangle-(N+1)(\N+2)L_{-1}|\tilde{B}_{F}\rangle\IEEEyessubnumber
\eea
where
\bea{l}
|\tilde{S}_{F}\rangle=\frac{1}{N\N}\ecd T'_{0}|S_{B}\rangle~,\\
|\tilde{A}_{F}\rangle=-\frac{(N+1)(\N-1)}{(N+\N+1)(N+2)\N}\ecd L_1|S_{B}\rangle+\frac{1}{N+\N+1}\ecd L_{-1}|B_{B}\rangle~,\IEEEyesnumber\\
|\tilde{B}_{F}\rangle=\frac{\N+2}{(N+2)(\N+1)}\ecd T'_0|B_{B}\rangle+i\frac{\N+2}{(\N+1)(N+\N+3)}\ead L_{-1}|B_{B}\rangle\\
~~~~~~~~~-i\frac{N+1}{(N+2)(N+\N+3)}\ead L_1|S_{B}\rangle~.
\eea
The result is that states $|\tilde{S}_{F}\rangle,~|\tilde{A}_{F}\rangle,~|\tilde{B}_{F}\rangle$ on-shell satisfy equations ($\ref{BRSTeom-F}$) and the theory of half-integer spins is invariant under the transformations
\bea{l}
\d_S |S_F\rangle=\frac{1}{N\N}\ecd T'_{0}|S_{B}\rangle~,\\
\d_S |A_F\rangle=-\frac{(N+1)(\N-1)}{(N+\N+1)(N+2)\N}\ecd L_1|S_{B}\rangle+\frac{1}{N+\N+1}\ecd L_{-1}|B_{B}\rangle~,\IEEEyesnumber\label{ds2}\\
\d_S |B_F\rangle=\frac{\N+2}{(N+2)(\N+1)}\ecd T'_0|B_{B}\rangle+i\frac{\N+2}{(\N+1)(N+\N+3)}\ead L_{-1}|B_{B}\rangle\\
~~~~~~~~~-i\frac{N+1}{(N+2)(N+\N+3)}\ead L_1|S_{B}\rangle~.
\eea

Transformations ($\ref{ds1},\ref{ds2}$) are exactly the on-shell
supersymmetry transformations for massless integer superspin
\cite{Kuz1}. Similar calculations can be done to demonstrate that
there is another transformation
\bea{l}
\delta|S_B\rangle=\ecd |S_F\rangle~,\\
\delta|B_B\rangle=\frac{(\N+2)^2}{\N+1}\ecd |B_F\rangle+i\frac{\N+2}{\N+1}\ead |A_F\rangle~,\IEEEyesnumber\label{susy3}\\
\hline\\
\delta|S_F\rangle=\frac{1}{(N+1)(\N+1)}\ea T'_0 |S_B\rangle-i\frac{\N}{(N+\N+1)(N+1)(\N+1)}\ec L_1 |S_B\rangle\\
~~~~~~~~~~~~+\frac{i}{(N+\N+1)N}\ec L_{-1} |B_B\rangle~,\\
\delta|A_F\rangle=-\frac{(N+2)\N}{(N+\N+3)(\N+1)}\ea L_1 |S_B\rangle+\frac{N+3}{N+\N+3}\ea L_{-1} |B_B\rangle~,\IEEEyesnumber\label{susy4}\\
\delta|B_F\rangle=\frac{1}{\N+3}\ea T'_0 |B_B\rangle~.
\eea
which corresponds to the on-shell
supersymmetry transformations for massless half-integer superspin.

\subsection{Off-shell formulation}
In this subsection we discuss a possibility to construct off-shell
supersymmetric Lagrangian formulation in the framework of the BRST
construction. First of all, one updates the information of the BRST
charge regarding supersymmetry. We have seen that for
$\mathcal{N}$=0, BRST can give us part of the off-shell
supersymmetric spectrum. However, we can see that in principle there
is room for the extra supersymmetric auxiliary components to appear.

In the case of half-ineger spin, the auxiliary component
$|\rho\rangle$ appeared in the lagrangian ($\ref{LF}$) as the
lagrange multiplier in order to generate the equation
$|\b\rangle=0$. But we could have seen the existance of
$|\rho\rangle$ as a redefinition of state $|A_F\rangle$ to
$|A_F\rangle+|\rho\rangle$ in ($\ref{0gh-F}$). The equation of
motion and the gauge transformation of $|A_F\rangle$ would remain
the same if $|\rho\rangle=0$ and $\delta_G|\rho\rangle=0$, exactly
as predicted by lagrangian ($\ref{LF}$).

For the integer spin case, the $\mathcal{N}$=0 BRST construction gives rise to one auxiliary component $|\gamma\rangle$. It is gauge invariant and has mass dimensions two. Therefore, all the bosonic auxiliary components in the various supersymmetric theories will reside in it and can be viewed as components of
\bea{l}
|\gamma\rangle\sim|A\rangle+|u\rangle+|v\rangle+|S\rangle+|P\rangle+|U\rangle
\eea
a deeper internal structure that has knowledge of supersymmetry. For consistency, they must be gauge invariant and vanish on-shell, i.e exactly as they behave in the supersymmetric theory. At the lagrangian description, the $\langle \gamma|\gamma\rangle$ in ($\ref{Lb}$) will be replaced by the sum of the diagonal terms
$\eta_A\langle A|A\rangle+\eta_u\langle u|u\rangle+\eta_v\langle v|v\rangle+\eta_S\langle S|S\rangle+\eta_P\langle P|P\rangle+\eta_U\langle U|U\rangle$.
The phases $\eta$ will have to be chosen such that they match their correct relative signs. For example in the integer superspin case the corresponding bosonic part of the lagrangian will be
\bea{ll}
\mathcal{L}_{B}=&\langle S_B|L_0|S_B\rangle-\langle S_B|\frac{1}{N+\N}L_{-1}L_1|S_B\rangle\\
&-\langle S_B|\frac{1}{N+\N}L_{-1}L_{-1}|B_B\rangle-\langle B_B|\frac{1}{N+\N+4}L_1L_1|S_B\rangle\\
&-2\langle B_B|\frac{N+\N+3}{N+\N+4}L_0|B_B\rangle+\langle B_B|\frac{1}{N+\N+4}L_{-1}L_1|B_B\rangle\\
&-\langle A|A\rangle+\langle u|u\rangle+\langle v|v\rangle-\langle S|S\rangle-\langle P|P\rangle-\langle U|U\rangle
\eea
and for the two half-integer superspins
\bea{ll}
\mathcal{L}^{\bot}_{B}=&\langle S_B|L_0|S_B\rangle-\langle S_B|\frac{1}{N+\N}L_{-1}L_1|S_B\rangle\\
&-\langle S_B|\frac{1}{N+\N}L_{-1}L_{-1}|B_B\rangle-\langle B_B|\frac{1}{N+\N+4}L_1L_1|S_B\rangle\\
&-2\langle B_B|\frac{N+\N+3}{N+\N+4}L_0|B_B\rangle+\langle B_B|\frac{1}{N+\N+4}L_{-1}L_1|B_B\rangle\\
&+\langle A|A\rangle+\langle u|u\rangle-\langle v|v\rangle-\langle S|S\rangle-\langle P|P\rangle+\langle U|U\rangle
\eea
\bea{ll}
\mathcal{L}^{\|}_{B}=&\langle S_B|L_0|S_B\rangle-\langle S_B|\frac{1}{N+\N}L_{-1}L_1|S_B\rangle\\
&-\langle S_B|\frac{1}{N+\N}L_{-1}L_{-1}|B_B\rangle-\langle B_B|\frac{1}{N+\N+4}L_1L_1|S_B\rangle\\
&-2\langle B_B|\frac{N+\N+3}{N+\N+4}L_0|B_B\rangle+\langle B_B|\frac{1}{N+\N+4}L_{-1}L_1|B_B\rangle\\
&+\langle A|A\rangle-\langle u|u\rangle-\langle v|v\rangle-\langle S|S\rangle+\langle P|P\rangle-\langle U|U\rangle~.
\eea
Also keep in mind that the value of the counting operators $N,~\bar{N}$ as they act on these states will be different, meaning that they have different index structure depending the system we describe. But this piece of information is encoded in the oscillators expansion of the Fock state.
\section[massive case]{Generalization for massive higher spin \\ supersymmetric theories}
Here we briefly describe a possible way to generalized the BRST
construction, described above, for massive supersymmetric higher
spin theories.

For the massive case, the set of constraints that define the
irreducible representations are modified in the following way:
\begin{itemize}
\item For half integer spin ($s+1/2$) we must have
\bea{l}
\pa^{\b\bd}\psi_{\b\a(s)\bd\ad(s-1)}=0,~ i\pa^{\b}{}_{\ad_{s+1}}\psi_{\b\a(s)\ad(s)}+m\bar{\psi}_{\a(s)\ad(s+1)}=0~.\IEEEyesnumber
\label{con.2}
\eea
\item For integer spin ($s$) we must have
\bea{l}
\pa^{\b\bd}h_{\b\a(s-1)\bd\ad(s-1)}=0~,~(\Box-m^2)h_{\a(s)\ad(s)}=0~.\IEEEyesnumber
\label{con.B2}
\eea
\end{itemize}
It is known that Lagrangians that describe the two these systems are
more complicate and involve a tower of real tensors of increasing
rank
$\{h,h_{a\ad},h_{\a(2)\ad(2)},...,h_{\a(s-2)\ad(s-2)},$ $h_{\a(s)\ad(s)}\}$
for the integer spin case
and a similar tower of tensors of increasing rank\\
$\{(\psi_{\a},\chi_{\a}),~(\psi_{\a(2)\ad},\chi_{\a(2)\ad}),\dots,(\psi_{\a(s-1)\ad(s-2)},\chi_{\a(s-1)\ad(s-2)}),~\psi_{\a(s)\ad(s-1)},~\psi_{\a(s+1)\ad(s)}\}$\\ for the half integer case. A physical way to understand why this is the case is to view the $2s+1$ states of spin $s$ as a collection of massless spin states with helicity $0$ up to helicity $s$ and similarly the $2s+2$ states of massive spin $s+1/2$ as a collection of massless spin states with helicity $1/2$ up to $s+1/2$.

To get the BRST description for the massive theory we have to follow similar steps as before, like converting all the above constraints into Fock space operators and calculate their algebra. Such a procedure could be done for integer spin, but for the half integer spin there is an obstacle. The Dirac equation, in this notation involves two states ($|\psi\rangle$ and $|\bar{\psi}\rangle$), therefore it can not be expressed as a single Fock operator acting on one Fock state. It looks like, we have to involve both subalgebras $\{T_0~,L_1~,L_{-1}~,L_0~,N~,\N\}$, $\{T'_0~,L_1~,L_{-1}~,L_0~,N~,\N\}$ and construct their direct sum. A possible way of doing that is to define a fermionic BRST charge in the following way
\bea{l}
Q\sim \begin{pmatrix} Q_F &  \\  & \bar{Q}_F \end{pmatrix},~
|\psi\rangle\sim \begin{pmatrix} |\psi_F\rangle  \\  |\bar{\psi}_F\rangle \end{pmatrix}~.\IEEEyesnumber
\eea
If that is the case, then the extension to the massive case would correspond to the existence of off-diagonal elements proportional to mass $m$.

Following this idea, we see that (\ref{con.2}) and its complex conjugate can be written in the following way
\bea{l}
\begin{pmatrix} \pa^{\b\bd} & 0 \\ 0 & \pa^{\b\bd} \end{pmatrix}\begin{pmatrix} \psi_{\b\a(s)\bd\ad(s-1)}  \\  \bar{\psi}_{\b\a(s-1)\bd\ad(s)}  \end{pmatrix}=0,~
\begin{pmatrix} i\pa^{\a_{s+1}}{}_{\ad_{s+1}} & m \\ m & i\pa_{\a_{s+1}}{}^{\ad_{s+1}} \end{pmatrix}\begin{pmatrix} \psi_{\a(s+1)\ad(s)}  \\  \bar{\psi}_{\a(s)\ad(s+1)}  \end{pmatrix}=0
\eea
and therefore we define,
\bea{l}
\hat{T}_0=\begin{pmatrix} T_0 & m \\ m & T'_0 \end{pmatrix},~\hat{L}_{0}=\begin{pmatrix} L_{0}-m^2 & 0 \\ 0 & L_{0}-m^2 \end{pmatrix}\\
\hat{L}_1=\begin{pmatrix} L_1 & 0 \\ 0 & L_1 \end{pmatrix},~\hat{L}_{-1}=\begin{pmatrix} L_{-1} & 0 \\ 0 & L_{-1} \end{pmatrix}
\eea

The rest of the procedure will follow the method developed earlier
for BRST construction in massive higher spin theories \cite{Buch4}.
It means that we have to get an extension of the algebra
($\ref{algebra}$) with mass terms playing the role of central
charges and therefore second class constraints. The conversion of
the second class constraints due to mass, to first class will be
done in the usual way by expanding our Fock space with more
oscillators that will lead to the presence of the extra states
required by the Lagrangian description. The exact construction will
be investigated in another letter.

\section{Problem of superfield BRST construction}
Here we briefly describe a possible way to formulate the BRST
construction for ${\cal N}=1$ supersymmetric higher spin theories in
superfield form and discuss the difficulties of such an approach.

The irreducible representations of Super-Poincare group have been are defined by the following constraints
\bea{l}
\D^{\b}H_{\b\a(s-1)\ad(s)}=0~,~\Dd^{\bd}H_{\a(s)\bd\ad(s-1)}=0~,~\D^{\g}\Dd^2\D_{\g}
H_{\a(s)\ad(s)}=0\IEEEyesnumber \eea for massless, half-integer
superspin where $H_{\a(s)\ad(s)}$ is a real bosonic superfield and
\bea{l}
\D^{\b}\Psi_{\b\a(s)\ad(s-1)}=0~,~\Dd^{\bd}\Psi_{\a(s+1)\bd\ad(s-1)}~,~\D^{\b}\Dd_{(\ad_{s+1}}\Psi_{\b\a(s)\ad(s))}=0\IEEEyesnumber
\eea
for massless, integer superspin. The extension of these for the massive irreps can be done by simply adding the mass term to the appropriate equations.
In the same spirit, we should define the corresponding Fock space
operators \bea{l} L_1=\D^{\b}a_{\b}~,~L_{2}=\Dd^{\bd}\ba_{\bd}~,~
L_{-2}=\D_{\a}a^{\dag}{}^{\a}~,~L_{-1}=\Dd_{\ad}\ba^{\dag}{}^{\ad}~,~\dots\IEEEyesnumber
\eea
calculate the algebra and attempt to build a BRST charge.
However, it is not obvious if the algebra closes or keeps
generating increasing powers of $\Box$ and whether it must be
interpreted as a nonlinear algebra or not. Furthermore, it is not
obvious how to choose the vacuum state such that, we generate
exactly the desired equations of motion. Also, we have to consider
the possibility to introduce matrix like operators in order to match
the component discussion.

The simplest gauge invariant theory in superspace is the massless
vector multiplet. The equation of motion is \bea{l} \D^\g\Dd^2\D_\g
H=0\IEEEyesnumber\label{S1} \eea and it is invariant under the
transformation
\bea{l}
\delta H=\D^2\bar{L}-\Dd^2 L~.\IEEEyesnumber
\eea
Therefore, the BRST charge must include the operators
$\D^2,~\Dd^2,~\D^\g\Dd^2\D_\g$ which have the algebra
\bea{l}
[\D^\g\Dd^2\D_\g,\D^2]=[\D^\g\Dd^2\D_\g,\Dd^2]=0,~[\D^2,\D^2]=-\frac{i}{2}\pa^{\a\ad}[\D_{\a},\Dd_{\ad}]~.\IEEEyesnumber
\eea
To close the algebra, without introducing operators with higher
rank or make it non-linear, we add to our list of operators
$\Delta=[\D^2,\D^2]$ and we attempt the construction of a BRST
charge
$Q=\eta D^\g\Dd^2\D_\g +\zeta\D^2+\xi\Dd^2+\rho[\D^2,\Dd^2]+c$.\\
In order $Q$ to be nilpotent we must satisfy the following:
\bea{l}
\eta^2=c^2=\rho^2=0~,\\
\{\eta,c\}=\{\zeta,c\}=\{\xi,c\}=\{\zeta,\xi\}=0~,\IEEEyesnumber\\
{[}\rho,\zeta ]=[\rho,\xi ]=0~,\\
\zeta\xi+\{\rho,c\}=0~.
\eea
Notice that $\zeta$ and $\xi$ do not have to be nilpotent and also $\rho$
has to commute with them. A solution of them is to have $\rho=-\zeta\xi P_{c}$ where $\{c,P_c\}=1$,
but we can not find a choice of vacuum that will generate equation (\ref{S1}).

An observation is that various superspace operators (like $\D^2$ and
$\Dd^2$) have by themselves the nilpotent property hence, they do not
require the existence of a nilpotent ghost in the
corresponding BRST charge. Another observation is that, unlike the
spacetime constructions where the nature of the constraints is
always the same, in superspace depending on the number of covariant
derivatives we will have even or odd parity operators therefore we
need to have both fermionic and bosonic ghost like oscillators.

The construction of a BRST charge for superspace, is unclear yet. A possible direction is to
look for a different but equivalent set of constraints to define the
irreducible representations in a way that allows a direct
application of the known methods. Another way, is to consider the possibility of
generalized charges where the kernel of $Q$ does not include the image of $Q$. This will relax the nilpotent condition to a
more general requirement for gauge invariance $Q\cdot Q'=0$, where
$Q$ is responsible for the equations of motion and $Q'$ for the
gauge transformations.
\section{Summary}

We have presented the BRST approach to Lagrangian description of 4D,
${\cal N}=1$ supersymmetric free massless higher spin theories
using the spinorial index notation. Generalization of the BRST
construction for the supersymmetric higher spin models was given and
the supersymmetric higher spin field Lagrangian was obtained.

We have demonstrated that the BRST description of massless free
higher spins in the 2-component spinor notation is preferable
because it circumvents the second class constraints related with the
trace and $\gamma$-trace conditions in the definition of higher
spins.

Furthermore, we have shown that the BRST procedure not only provides all
the necessary auxiliary fields for the higher spin description but also give
rise to extra auxiliary components that play the role of off-shell supersymmetric auxiliary components. In the other words,
one can expect that the BRST approach will give a possibility of
complete off-shell Lagrangian formulation of higher spin
supersymmetric models. A hint for that is lagrangian $(\ref{LF})$
which matches exactly the fermionic part of the corresponding superspace lagrangians
that describe integer and half-integer superspin irreducible representations of
the super-Poincar\'{e} group.

Finally, we have illustrated how, on-shell supersymmetry
transformations emerge just because, the algebra of the bosonic
constraints is included in the algebra of the fermionic constraints.

All the above results motivates us to carry out a similar analysis
for the massive higher spin supersymmetric models and also for
superspace formulated theories. In the first case, it seems that we
have to expend our Fock space, so that for every state we introduce
its complex conjugate. That forces us to redefine the various
operators of the massless theory in terms of matrices acting on a complex conjugated enhanced Fock vector. Afterwards, we have to
calculate the algebra of these objects and check a) if the usual
BRST procedure can apply and b) if there is a vacuum state that will
give the desired equations of motion. Also, we can apply the methods
developed earlier for BRST construction in massive higher spin
theories \cite{Buch4}. For the superspace case, the picture is not
so clear regarding the closure of the algebra of the constraints and
its peculiarities like the fact that it already includes nilpotent
objects. We plan to study all these aspects in the forthcoming
works.

When all the above are understood and we have a good understanding of the
free supersymmetric theories viewed from the BRST point of view, we can start asking questions about interactions. This will correspond to deforming the BRST charge by adding vertex operators that result into interactions among the various Fock space vectors. This will be an even more challenging problem because not only we will have to face the difficulties of interacting higher spins but we want to do it in a supersymmetric way.

\section*{Acknowledgements}
The authors are very grateful to V.A. Krykhtin and W.D. Linch for
helpful discussions. K.K. is thankful to Center of Theoretical
Physics at Tomsk State Pedagogical University for warm hospitality.
The work of I.L.B was supported by the RFBR grant, project No
15-02-03594 and by the Ministry of Education and Science of Russian
Federation, project 3.967.2014/K. The work of K.K was supported by
RFBR grant, project No 15-32-50086 and by P201/12/G028 of the Grant
agency of the Czech republic.

\newpage
\appendix
 \renewcommand{\theequation}{A-\arabic{equation}}
  \setcounter{equation}{0}  
\section{Two-component Spinor Notation}
Consider the 4D Minkowski, Clifford algebra
\bea{l}
\{e_m,e_n\}=2\eta_{mn}~.\IEEEyesnumber
\eea
It is well know, that the objects
\bea{l}
J_{mn}=-\frac{i}{4}[e_m,e_n],~P_m=e_m\IEEEyesnumber
\eea
satisfy the Poincar\'{e} algebra and therefore the representations of the Poincar\'{e} algebra
can be generated by the representations of the Clifford algebra. Also, it is well known
that in 4D we can have Weyl spinors, thus the smallest representations of Clifford algebra are the left $\psi_{\a}$  and right $\chi^{\ad}$ spinors. Under the action of the Poincare algebra they transform in the following way:
\bea{l}
[J_{mn},\psi_{\a}]=i(\sigma_{mn})_{\a}{}^{\b}\psi_{\b}~,~[J_{mn},\chi^{\ad}]=i(\bar{\sigma}_{mn})^{\ad}{}_{\bd}\chi^{\bd}\IEEEyesnumber
\eea
where
\bea{l}
(\sigma_{mn})_{\a}{}^{\b}=\frac{1}{4}(\sigma_m)_{\a\ad}(\bar{\sigma}_n)^{\ad\b}-\frac{1}{4}(\sigma_n)_{\a\ad}(\bar{\sigma}_m)^{\ad\b}~,\IEEEyesnumber\\
(\bar{\sigma}_{mn})^{\ad}{}_{\bd}=\frac{1}{4}(\bar{\sigma}_m)^{\ad\a}(\sigma_n)_{\a\bd}-\frac{1}{4}(\bar{\sigma}_n)^{\ad\a}(\sigma_m)_{\a\bd}~.
\eea
The $\sigma_m=(1,\vec{\sigma})$ and $\bar{\sigma}_m=(-1,\vec{\sigma})$\footnote{$\vec{\sigma}$ are the three Pauli matrices} satisfy a list of very useful properties
\bea{l}
(\sigma_m)_{\a\ad}(\bar{\sigma}_n)^{\ad\b}+(\sigma_n)_{\a\ad}(\bar{\sigma}_m)^{\ad\b}=2\eta_{mn}\d_{\a}{}^{\b}~,\\
(\bar{\sigma_m})^{\ad\a}(\sigma_n)_{\a\bd}+(\bar{\sigma}_n)^{\ad\a}(\sigma_m)_{\a\bd}=2\eta_{mn}\d^{\ad}{}_{\bd}~,\\
(\sigma^m)_{\a\ad}(\bar{\sigma}_m)^{\bd\b}=2\d_{\a}{}^{\b}\d_{\ad}{}^{\bd}~,\\
\tfrac{1}{2}\e^{klmn}\sigma_{mn}=-i\sigma^{kl}~~,~~\tfrac{1}{2}\e^{klmn}\bar{\sigma}_{mn}=i\bar{\sigma}^{kl}\IEEEyesnumber\label{A1}~,\\
\sigma^k\bar{\sigma}^{ln}=\tfrac{1}{2}\eta^{lk}\sigma^n-\tfrac{1}{2}\eta^{nk}\sigma^l-\tfrac{i}{2}\epsilon^{klnm}\sigma_m~,\\
\bar{\sigma}^k\sigma^{ln}=\tfrac{1}{2}\eta^{lk}\bar{\sigma}^n-\tfrac{1}{2}\eta^{nk}\bar{\sigma}^l+\tfrac{i}{2}\epsilon^{klnm}\bar{\sigma}_m~,\\
\sigma^{ln}\sigma^k=-\tfrac{1}{2}\eta^{lk}\sigma^n+\tfrac{1}{2}\eta^{nk}\sigma^l-\tfrac{i}{2}\epsilon^{klnm}\sigma_m~,\\
\bar{\sigma}^{ln}\bar{\sigma}^k=-\tfrac{1}{2}\eta^{lk}\bar{\sigma}^n+\tfrac{1}{2}\eta^{nk}\bar{\sigma}^l+\tfrac{i}{2}\epsilon^{klnm}\bar{\sigma}_m~.
\eea

The $(\sigma_m)_{\a\ad},~(\bar{\sigma}_m)^{\ad\a}$ are the only objects that have all three kinds of indices. For this reason they are used for converting vector indices to left-right indices and vice versa. For example:
\bea{l}
A_m=(\bar{\sigma}_m)^{\ad\a}A_{\a\ad}~,~A_{\a\ad}=\frac{1}{2}(\sigma^m)_{\a\ad}A_m~.\IEEEyesnumber
\eea
An example of that would be the partial derivative, $\pa_m$. So let's define
\bea{l}
\pa_{\a\ad}=(\sigma^m)_{\a\ad}\pa_m~.\IEEEyesnumber
\eea
We can show that it has the properties
\bea{l}
\pa^{\a\ad}\pa_{\a\bd}=\d^{\ad}{}_{\bd}\Box ~,~\pa^{\ad\a}\pa_{\a\bd}=\d^{\ad}{}_{\bd}\Box~,\\
\pa^{\a}{}_{\bd}\pa_{\b}{}^{\ad}-\pa^{\a\ad}\pa_{\b\bd}=\d^{\a}{}_{\b}\d^{\ad}{}_{\bd}~.\Box\IEEEyesnumber
\eea
The conversion of vector indices to spinorial ones, doesn't work just for vectors but for higher rank tensors as well. For example consider the case of a rank two tensor
\bea{l}
A_{mn}=(\bar{\sigma}_m)^{\ad\a}(\bar{\sigma}_n)^{\bd\b}A_{\a\b\ad\bd}\IEEEyesnumber
\eea
$A_{\a\b\ad\bd}$ can be further decomposed by symmetrizing and anti-symmetrizing the undotted and dotted pair of indices.
\bea{l}
A_{\a\b\ad\bd}=A^{(S,S)}_{\a\b\ad\bd}+C_{\a\b}A^{(A,S)}_{\ad\bd}+C_{\ad\bd}A^{(S,A)}_{\a\b}+C_{\a\b}C_{\ad\bd}A^{(A,A)}\IEEEyesnumber\label{A2}
\eea
and we get that
\bea{l}
A^{(S,S)}_{\a\b\ad\bd}=\tfrac{1}{16}(\sigma^m)_{(\a(\ad}(\sigma^n)_{\b)\bd)}A_{mn}~,\\
A^{(A,S)}_{\ad\bd}=-\tfrac{1}{4}(\bar{\sigma}^{mn})_{\ad\bd}A_{mn}~,\\
A^{(S,A)}_{\a\b}=\tfrac{1}{4}(\sigma^{mn})_{\a\b}A_{mn}~,\IEEEyesnumber\label{A3}\\
A^{(A,A)}=\tfrac{1}{8}\eta^{mn}A_{mn}~.
\eea
From the above we can see that for a rank two antisymmetric tensor (like the generators of the Lorentz group,~$\mathcal{J}_{mn}$) the completely symmetric and the scalar terms vanish ($A^{(S,S)}_{\a\b\ad\bd}=0,~A^{(A,A)}=0$) and we get
\bea{l}
\mathcal{J}_{mn}=2(\sigma_{mn})^{\a\b}\mathcal{J}_{\a\b}-2(\bar{\sigma}_{mn})^{\ad\bd}\mathcal{J}_{\ad\bd}\IEEEyesnumber
\eea

The $C_{\a\b}$ and $C_{\ad\bd}$ are antisymmetric objects defined in the following way
\bea{l}
C^{\a\b}=\left(\begin{array}{cc}0 & i\\ -i & 0\end{array}\right)=C^{\ad\bd}~,~
C_{\a\b}=\left(\begin{array}{cc}0 & -i\\ i & 0\end{array}\right)=C_{\ad\bd}\IEEEyesnumber
\eea
and they are also used as a metric to raise and lower the spinorial indices.
They have the property
\bea{l}
C^{\a\b}C_{\g\r}=\d^{\a}{}_{\g}\d^{\b}{}_{\r}-\d^{\a}{}_{\r}\d^{\b}{}_{\g}~,\\
C^{\ad\bd}C_{\gd\rd}=\d^{\ad}{}_{\gd}\d^{\bd}{}_{\rd}-\d^{\ad}{}_{\rd}\d^{\bd}{}_{\gd}~.\IEEEyesnumber
\eea

For the description of higher spins we will use fields $\phi$ with $k$ number of undotted indices and $l$ number of dotted indices and have the property that are independently symmetrized in both of them. We will denote that by writing $\phi_{\a(k)\ad(l)}$. Then from (\ref{A1},~\ref{A2},~\ref{A3}) follows that the trace and $\gamma$-trace of these fields vanish identically. That is very useful because in this formulation of higher spins, the trace constraints that usually introduce second class constraints in the BRST approach are not present any more.


%
%
\footnotesize{

\end{document}